\documentclass[preprintnumbers, floatfix, preprintnumbers, letterpaper, superscriptaddress,nofootinbib]{revtex4}
\pdfoutput=1
\usepackage{graphicx}
\usepackage{microtype}
\usepackage{amsmath}
\usepackage{amssymb}
\usepackage{subfigure}
\usepackage{hyperref}
\usepackage{url}
\usepackage{xcolor}
\usepackage{color}
\usepackage{mathrsfs}
\usepackage{calrsfs}
\usepackage{amsfonts}
\usepackage{latexsym}
\usepackage{ragged2e}
\usepackage{epsfig}
\usepackage{textcomp}
\usepackage{phaistos}
\usepackage[utf8]{inputenc}
\makeatletter
\renewcommand\@makefnmark{\hbox{\@textsuperscript{\normalfont\color{purple}\@thefnmark}}}
\renewcommand\@makefntext[1]{%
  \parindent 1em\noindent
            \hb@xt@1.8em{%
                \hss\@textsuperscript{\normalfont\@thefnmark}}#1}
\makeatother

\usepackage{caption}
\DeclareCaptionJustification{justified}{\leftskip=0pt \rightskip=0pt \parfillskip=0pt plus 1fil}
\definecolor{vividviolet}{rgb}{0.62, 0.0, 1.0}
\definecolor{amaranth}{rgb}{0.9, 0.17, 0.31}
\definecolor{palatinateblue}{rgb}{0.15, 0.23, 0.89}
\definecolor{brightpink}{rgb}{1.0, 0.0, 0.5}
\definecolor{cornflowerblue}{rgb}{0.39, 0.58, 0.93}
\definecolor{deepcarminepink}{rgb}{0.94, 0.19, 0.22}
\definecolor{radicalred}{rgb}{1.0, 0.21, 0.37}

\hypersetup{ linktoc=all,
    colorlinks, linkcolor={palatinateblue},
    citecolor={brightpink}, urlcolor={amaranth}
}
\newcommand{\changeurlcolor}[1]{\hypersetup{urlcolor=#1}}
\graphicspath{{Images/}}

\def\sideremark#1{\ifvmode\leavevmode\fi\vadjust{\vbox to0pt{\vss
 \hbox to 0pt{\hskip\hsize\hskip1em
 \vbox{\hsize1.5cm\tiny\raggedright\pretolerance10000
 \noindent #1\hfill}\hss}\vbox to8pt{\vfil}\vss}}}%
\begin{document}

\title{Regular Black Hole Interior Spacetime Supported by Three-Form Field}

\author{Mariam Bouhmadi-L\'opez}
\email{mariam.bouhmadi@ehu.eus}
\affiliation{Department of Physics, University of the Basque Country UPV/EHU, Bilbao 48080, Spain}
\affiliation{IKERBASQUE, Basque Foundation for Science, Bilbao 48011, Spain}

\author{Che-Yu Chen}
\email{b97202056@gmail.com}
\affiliation{Department of Physics and Center for Theoretical Sciences, National Taiwan University, Taipei 10617, Taiwan}
\affiliation{Leung Center for Cosmology and Particle Astrophysics, National Taiwan University, Taipei 10617, Taiwan}
\affiliation{Institute of Physics, Academia Sinica, Taipei 11529, Taiwan}

\author{Xiao Yan \surname{Chew}}
\email{xychew998@gmail.com}
\affiliation{Department of Physics Education, Pusan National University, Busan 46241, Republic of Korea}
\affiliation{Research Center for Dielectric and Advanced Matter Physics, Pusan National University, Busan 46241, Republic of Korea}

\author{Yen Chin \surname{Ong}}
\email{ycong@yzu.edu.cn}
\affiliation{Center for Gravitation and Cosmology, College of Physical Science and Technology, Yangzhou University, \\180 Siwangting Road, Yangzhou City, Jiangsu Province  225002, China}
\affiliation{School of Aeronautics and Astronautics, Shanghai Jiao Tong University, Shanghai 200240, China}

\author{Dong-han Yeom}
\email{innocent.yeom@gmail.com}
\affiliation{Department of Physics Education, Pusan National University, Busan 46241, Republic of Korea}
\affiliation{Research Center for Dielectric and Advanced Matter Physics, Pusan National University, Busan 46241, Republic of Korea}

\begin{abstract}
In this paper, we show that a minimally coupled 3-form endowed with a proper potential can support a regular black hole interior. By choosing an appropriate form for the metric function representing the radius of the 2-sphere, we solve for the 3-form field and its potential. Using the obtained solution, we construct an interior black hole spacetime which is everywhere regular. The singularity is replaced with a Nariai-type spacetime, whose topology is $\text{dS}_2 \times \text{S}^2$, in which the radius of the 2-sphere is constant. So long as the interior continues to expand indefinitely, the geometry becomes essentially compactified. The 2-dimensional de Sitter geometry appears despite the negative potential of the 3-form field. Such a dynamical compactification could shed some light on the origin of de Sitter geometry of our Universe, exacerbated by the Swampland conjecture. In addition, we show that the spacetime is geodesically complete. The geometry is singularity-free due to the violation of the null energy condition.
\end{abstract}

\maketitle

\section{Introduction: Regular Black Holes}

Singularities are commonplace in general relativity, as established by the singularity theorems \cite{1,1-2,jose}. In particular, singularities tend to form during gravitational collapse and thus black holes harbor singularities. For example, a Schwarzschild black hole has a spacelike singularity, which lies in the future of an infalling observer. Reissner-Nordstr\"om black holes and Kerr black holes, on the other hand, possess timelike singularities. Curvature (as measured by, e.g., the Kretschmann invariant) becomes increasingly large as one approaches the singularity, which indicates that new physics could come into play near it. That is, quantum gravitational effect is usually expected to cure spacetime singularities. 

We do not yet have a fully working theory of quantum gravity, but black holes without singularity -- ``regular black holes'' -- have been studied in great detail, mostly as phenomenological models. In fact, it is interesting to investigate how regular black hole solutions can be obtained even at the classical level, in the presence of some matter fields. Ad hoc though it may be to consider somewhat exotic fields, such as 1-form potential within nonlinear electrodynamics, this at least gives us some comfort that in the regime of quantum gravity, a plethora of new fields that could be excited at higher energy scales would indeed couple to gravity to ``regularize'' black hole singularities, albeit most likely in a more complicated manner.

The first regular exact black hole solution in general relativity is a charged black hole supported by nonlinear electrodynamics \cite{9911046}. Its metric is given by
\begin{equation}
ds^2 = -\left[1-\frac{2mr^2}{(r^2+q^2)^{3/2}} + \frac{q^2r^2}{(r^2+q^2)^2}\right] dt^2 + \left[1-\frac{2mr^2}{(r^2+q^2)^{3/2}} + \frac{q^2r^2}{(r^2+q^2)^2}\right]^{-1} dr^2 + r^2d \Omega^2,\label{nedmetric}
\end{equation}
where $d\Omega^2$ is the standard metric on a 2-sphere, and $m,q$ are related to the mass and charge, respectively. This solution behaves like a Reissner-Nordstr\"om black hole asymptotically as $r \to \infty$. Like Reissner-Nordstr\"om, this black hole also has two horizons. The presence of the inner (Cauchy) horizon is the reason why singularity theorem does not apply -- the theorem requires the underlying spacetime to be globally hyperbolic (of course, even when the premise of the theorem is false, it does not mean that the spacetime is guaranteed to be singularity free, as shown by the Reissner-Nordstr\"om black hole). The interior spacetime below the inner horizon is static, with a regular origin at $r=0$. This is, in fact, a charged version of the well-known Bardeen solution \cite{b1,b2}. 

How about higher form fields\footnote{As far as \emph{exterior} geometry is concerned, due to the no-hair theorem, static black holes in $n$-dimensional spacetime (either asymptotically flat, de Sitter or anti-de Sitter) cannot admit $p$-form charge for $p \geqslant 4$ (in our terminology), at least when there is no nontrivial potential \cite{1007.3847,1101.1121}. }? Let us consider a 3-form field $\textbf{A}$ (by which we mean that the potential is a 3-form, which means the analogues ``Faraday tensor'' or field strength, $\textbf{F}=d\textbf{A}$, is a 4-form). In the absence of a nontrivial potential $V$ in the Lagrangian (not to be confused with the 3-form potential $\textbf{A}$), it behaves like a cosmological constant{\footnote{It is true that the presence of a potential breaks the gauge invariance present initially in the theory but likewise this happens for the inflaton field; i.e. the kinetic term is invariant under a translation as long as the potential is constant, however, as soon as the scalar field potential is switched on, such an invariance is gone. The same happens with a Proca field.}} \cite{DUFF1980179}. In fact, we consider such a form field because it is well-motivated, and its roles in cosmology have been widely investigated. Indeed those fields have been proven to be useful in string cosmology \cite{Copeland:1994km,Lukas:1996iq}, the pre-big-bang scenario  \cite{Gasperini:1998bm} and quantum cosmology \cite{BBG}. More recently, they have been widely analysed in the paradigm of inflation \cite{GK09,Koivisto:2009sd,Germani:2009gg,Koivisto:2009fb,Koivisto:2009ew,DeFelice:2012jt,DeFelice:2012wy}. Likewise, primordial non-gaussianity induced by one or several 3-forms has been analysed in \citep{Mulryne:2012ax,Kumar:2014oka,Kumar:2016tdn}. Most importantly, this scenario has been successfully fitted to the current Planck data. The 3-forms have been proven to be proliferous: they are also able to describe the late-time acceleration of the Universe \cite{Koivisto:2009fb,Koivisto:2009ew}, which does not come as a big surprise, as a potential free 3-form mimics a cosmological constant \cite{DUFF1980179}. What is truly innovative about 3-forms in what refers to the late-time Universe is that they can mimic a phantom dark energy with a standard kinetic term as long as the 3-form potential is positive and a decreasing function of the contracted square of the 3-form \cite{3formCovilha1,3formCovilha2}.

Given that $p$-forms are ubiquitous in string theory, which is one of the candidates for quantum gravity, it is interesting to investigate whether 3-form can help to remove black hole singularities, in addition to its utility in cosmology. Indeed, axionic field $H_{\mu\nu\rho}$, which is the field strength of a 2-form field, can regularize flat (i.e. of planar or toral topology) locally asymptotically anti-de Sitter black holes \cite{1202.4458}. Previously, it was shown that axion, without nontrivial potentials, does not affect the exterior metric of an asymptotically flat black hole spacetime. Furthermore, the field strength vanishes in that case \cite{prl1988}. See also Ref.~\cite{Cisterna:2020rkc} in which black holes in higher dimensions with $p$-form fields are investigated.

For the case of 3-form, we find that with a certain choice of the potential $V$, which is not that complicated, 3-form field can give rise to an interior solution which is regular. However, unlike the aforementioned charged Bardeen black hole, our solution has only one horizon, so that the interior is a cosmological spacetime with time extending eternally into the infinite future, instead of a static ``core''. Even more surprisingly, as we will discuss in more detail in the Discussion section, its topology is $\text{dS}_2 \times \text{S}^2$, with the size of the 2-sphere being constant as the de Sitter part expands, effectively giving rise to some sort of compactification. This is different from the Frolov-Markov-Mukhanov model in which a Schwarzschild interior transits into a 4-dimensional de Sitter spacetime \cite{FMM1, FMM2}. That is to say, in our case the singularity is replaced by a Nariai-type universe \cite{Nariai1,Nariai2}. Similar regular black holes can also be formulated in loop quantum gravity, specifically in the so-called ``$\bar\mu$-type scheme'' \cite{Bohmer:2007wi,Boehmer:2008fz,Brannlund:2008iw,Dadhich:2015ora}. In this paper, we show that such a resolution of black hole singularity is attainable at the classical level, supported by 3-form fields. Furthermore, we hope that this toy model could hint at possible singularity avoidance in a realistic gravitational collapse, with 3-form fields activated by sufficiently large curvature, as we will explain. We will also show that the spacetime is geodesically complete and the reason why the singularity is avoided is due to the violation of the null energy condition.

Our manuscript is structured as follows: In Sec. (\ref{2}) we briefly review how some black hole interior geometries can be interpreted as a cosmological spacetime, taking Schwarzschild interior as an example. Then, in Sec. (\ref{3}) we compute the equations of motion for a 3-form minimally coupled to Einstein's gravity. We then study the solutions in detail in Sec. (\ref{4}). Finally we conclude in Sec. (\ref{5}) with some discussions. 

\section{Black hole interior and Kantowski-Sachs spacetime}\label{2}
Inside the event horizon, the spacetime of a static and spherically symmetric black hole can be described by the Kantowski-Sachs anisotropic cosmology \cite{Kantowski:1966te,Doran:2006dq}:
\begin{equation}
ds^2=-dt^2+a(t)^2d\mathcal{R}^2+r_s^2b(t)^2d\Omega_2^2\,,\label{KSmetric}
\end{equation}
where $a(t)$ and $b(t)$ are dimensionless scale factors. We introduce a parameter $r_s$, which has a length dimension, to denote the radius of the event horizon. For example, the interior metric of the Schwarzschild spacetime
\begin{equation}
ds_S^2=-\left(\frac{r_s}{r}-1\right)^{-1}dr^2+\left(\frac{r_s}{r}-1\right)d\mathcal{R}^2+r^2d\Omega_2^2
\end{equation}
can be rewritten in the form of the metric \eqref{KSmetric} after redefining the timelike variable $r\rightarrow r(t)$. For the Schwarzschild metric, the exact expressions of the scale factors can be obtained by defining a new time variable $\eta(t)$, such that \cite{Doran:2006dq}
\begin{equation}
\frac{t}{r_s}=\eta+\sin\eta\cos\eta\,,
\end{equation}
and the scale factors can be written as
\begin{equation}
a(t)=\tan\eta(t)\,,\qquad b(t)=\cos^2\eta(t)\,.
\end{equation}
Note that the two scale factors satisfy
\begin{equation}
b=\frac{1}{1+a^2}\,,
\end{equation}
and this relation is independent of the choice of the timelike variable.

Near the event horizon where $t/r_s\rightarrow0$, the scale factors of the Schwarzschild interor can be approximated as
\begin{equation}
\textrm{Event horizon:}\quad a(t)\approx\frac{t}{2r_s}\,,\qquad b(t)\approx 1-\frac{t^2}{4r_s^2}\,.\label{horizonlim}
\end{equation}
On the other hand, the singularity takes place when $t/r_s\rightarrow\pi/2$. At the singularity, the scale factors behave as
\begin{equation}
\textrm{Singularity:}\quad b(t)\approx\frac{1}{a(t)^2}\approx\left[\frac{3}{2}\left(\frac{\pi}{2}-\frac{t}{r_s}\right)\right]^{2/3}\,.
\end{equation}
The motivation of this work is to construct a regular black hole interior geometry by including 3-form fields with suitable potentials. We are interested in those solutions that can recover the Schwarzschild interior near the horizon ($t/r_s\rightarrow0$), while deviate from it when $t/r_s$ increases in a way that the singularity is replaced with a regular geometry. We want to recover the Schwarzschild interior near the horizon since we envision that in a more realistic model in which a black hole is formed in a gravitational collapse, 3-form fields, and possibly higher form fields in string theory only get activated at sufficiently large energy scales, i.e., when curvature is large enough. For a sufficiently massive black hole, the curvature at the horizon can be very small, so it makes sense to require such a boundary condition.

\section{Equations of motion}\label{3}
We consider a 3-form field $A_{\mu\nu\rho}$ minimally coupled to Einstein's general relativity (GR). The action can be written as
\begin{equation}
S=\int d^4x\sqrt{-g}\left[\frac{R}{2\kappa}-\frac{1}{48}\bold{F}^2-V\left(\bold{A}^2\right)\right]\,,\label{threeformaction}
\end{equation}
where $\kappa=8\pi G$ is the gravitational constant, and the field strength $F_{\mu\nu\rho\sigma}$ is defined by
\begin{align}
F_{\mu\nu\rho\sigma}&=4\nabla_{[\mu}A_{\nu\rho\sigma]}\nonumber\\
&=\nabla_\mu A_{\nu\rho\sigma}-\nabla_\sigma A_{\mu\nu\rho}+\nabla_\rho A_{\sigma\mu\nu}-\nabla_\nu A_{\rho\sigma\mu}\,.\label{strengthf}
\end{align}
In the action \eqref{threeformaction}, the scalar invariants $\bold{F}^2$ and $\bold{A}^2$ are defined by $\bold{F}^2\equiv F_{\mu\nu\rho\sigma}F^{\mu\nu\rho\sigma}$ and $\bold{A}^2\equiv A_{\mu\nu\rho}A^{\mu\nu\rho}$, respectively. Furthermore, $R$ is the Ricci scalar of the spacetime and the 3-form field is subject to a potential $V(\bold{A}^2)$. The Einstein equation can be obtained by varying the action \eqref{threeformaction} with respect to the metric $g_{\mu\nu}$:
\begin{equation}
R_{\mu\nu}-\frac{1}{2}g_{\mu\nu}R=\kappa T_{\mu\nu}\,,
\end{equation}
where $R_{\mu\nu}$ and $T_{\mu\nu}$ are the Ricci tensor and the energy-momentum tensor of the 3-form, respectively. The latter can be written explicitly as
\begin{equation}
T_{\mu\nu}=\frac{1}{6}F_{\mu\alpha\beta\gamma}{F_{\nu}}^{\alpha\beta\gamma}+6\frac{\partial V}{\partial\bold{A}^2}A_{\mu\alpha\beta}{A_\nu}^{\alpha\beta}-g_{\mu\nu}\left[\frac{1}{48}\bold{F}^2+V\left(\bold{A}^2\right)\right]\,.\label{energymomenttensor}
\end{equation}

In the Kantowski-Sachs spacetime, whose metric is given by Eq.~\eqref{KSmetric}, the non-vanishing components of the form field can be expressed as
\begin{equation}
A_{R\theta\phi}=r_s^2a(t)b(t)^2\chi(t)E_{R\theta\phi}\,, \quad F_{tR\theta\phi}=\frac{d}{dt}\left[r_s^2a(t)b(t)^2\chi(t)\right]E_{R\theta\phi}\,,
\end{equation}
where $E_{R\theta\phi}$ is the Levi-Civita tensor defined on a ($\mathbb{R}\times \text{S}^2$) spatial metric $ds_h^2=dR^2+d\Omega_2^2$, and $\chi(t)$ quantifies the dynamics of the form field. We then have
\begin{equation}
\bold{A}^2=6\chi^2\,,\qquad \bold{F}^2=-24\left(\dot\chi+H_a\chi+2H_b\chi\right)^2\,,
\end{equation}
where $H_a\equiv\dot{a}/a$ and $H_b\equiv\dot{b}/b$ are the Hubble parameters associated with the two scale factors, with dot denoting the derivatives with respect to $t$. In the Kantowski-Sachs spacetime, the equations of motion can be obtained as follows:
\begin{align}
&H_a^2+\dot{H}_a-2H_b^2-\dot{H}_b+H_aH_b-\frac{1}{r_s^2b^2}=0\,,\label{eqc}\\
&\ddot{\chi}+\left(\dot{H}_a+2\dot{H}_b\right)\chi+\left(H_a+2H_b\right)\dot\chi+\frac{dV}{d\chi}=0\,,\label{eqchi}\\
&\frac{1}{\kappa}\left(3H_b^2+2\dot{H}_b+\frac{1}{r_s^2b^2}\right)
=\chi\ddot{\chi}+\frac{\dot\chi^2}{2}+2\chi\dot\chi\left(H_a+2H_b\right)+\chi^2\left(\frac{H_a^2}{2}+2H_b^2+\dot{H}_a+2\dot{H}_b+2H_aH_b\right)+V(\chi)\,.\label{eqa}
\end{align}
In addition, the Hamiltonian constraint of the system is given by
\begin{equation}
H_b^2+2H_aH_b+\frac{1}{r_s^2b^2}=\frac{\kappa}{2}\left(\dot\chi+H_a\chi+2H_b\chi\right)^2+\kappa V(\chi)\,.\label{constraint}
\end{equation}
It should be emphasized that, unlike the standard scalar field which is minimally coupled to gravity, the kinetic term of the 3-form (see the right-hand side of Eq.~\eqref{constraint}) contains the contributions of $H_a$ and $H_b$.

\section{Regular black hole supported by 3-form}\label{4}
In the context of GR minimally coupled with 3-form fields, the intuitive approach to obtain black hole solutions is to start with assuming some particular potentials $V(\chi)$, then solve the field equations to get the solutions. This method has been utilized in Ref.~\cite{Barros:2020ghz}, to obtain spherically symmetric solutions. Some of these solutions can be interpreted as black holes by identifying the existence of event horizons (wormhole solutions were also found in \cite{1806.10488}). The others can be interpreted as naked singularities in which there is no event horizon in the spacetime. However, the analysis regarding the black hole solutions conducted in Ref.~\cite{Barros:2020ghz} only focuses on the exterior region of the spacetime. It is not clear how such black hole solutions behave inside the horizons. Since we are interested in formulating regular black hole solutions in this theory, we will mainly focus on the spacetime inside the event horizon, which, as we have mentioned, can be described by the Kantowski-Sachs metric \eqref{KSmetric}.

To construct regular black hole solutions, we will not assume a particular 3-form potential at the beginning, as apposed to the method used in Ref.~\cite{Barros:2020ghz}. Instead, we will consider a particular choice of the scale factor $b(t)$, such that instead of going to zero at a finite $t$ (the singularity in the Schwarzschild black hole takes place at $t/r_s=\pi/2$), it has a non-zero minimum value $b_m$ satisfying $1>b_m>0$. Using Eqs.~\eqref{eqc}, \eqref{eqchi}, \eqref{eqa}, and inserting proper initial conditions near the event horizon where $b\approx 1$, we can solve $H_a(t)$, $\chi(t)$, $V(t)$ numerically. After obtaining the solutions, one has to check whether the constraint equation \eqref{constraint} is satisfied.

The scale factor $b(t)$ of our interest, which could appear in regular black hole models, is
\begin{equation}
b(t)=\frac{1}{(x+1)^2}\left\{x+\textrm{exp}\left[-\frac{\left(x+1\right)t^2}{8r_s^2}+c_4(x)\left(\frac{t}{r_s}\right)^{4}\right]\right\}^2\,,\label{b1}
\end{equation}
where $c_4(x)<0$ is a constant coefficient. The inclusion of this coefficient is to justify the initial conditions, which will be presented later. The parameter $x>0$ is dimensionless and it directly determines $b_m$ as follows 
\begin{equation}
b_m=\left(\frac{x}{1+x}\right)^2\,.
\end{equation}
Note that the scale factor $b(t)$ given by Eq.~\eqref{b1} satisfies $b(t)\approx 1-t^2/4r_s^2$ when $t/r_s\rightarrow0$. Therefore, the scale factor $b(t)$ reduces to the Schwarzschild counterpart near the horizon (see Eq.~\eqref{horizonlim}). From now on, we will rescale the timelike coordinate as $t/r_s \mapsto t$, for the sake of simplicity.

As we have mentioned, the coefficient $c_4(x)$ is related to the justification of the initial conditions. First, we assume that at the initial point $t=t_i$ near the horizon, where $t_i\ll 1$, the 3-form field $\chi(t)$ and its derivative is given by
\begin{equation}
\kappa\chi(t_i)=t_i^3\,,\qquad \kappa\dot{\chi}(t_i)=3t_i^2\,.
\end{equation}   
With this assumption, the second and the third terms of Eq.~\eqref{eqchi} are of the order of $t_i$ near the horizon, provided that the leading order of $H_a(t_i)$ is $1/t_i$ (see Eq.~\eqref{horizonlim}). In order to ensure the smallness of the contributions from the 3-form field near the horizon, we have to further ensure that generically, the first and the last terms in Eq.~\eqref{eqchi}, that is, $\ddot{\chi}$ and $dV/d\chi$, are small near the horizon. Given that the kinetic terms of the right-hand side of Eqs.~\eqref{eqa} and \eqref{constraint} are generically of the order of $t_i^4$ near the horizon, we find that it is necessary to take into account the series expansion of $H_a(t)$ near $t_i$, up to the third order term. Therefore, the initial condition for $H_a(t_i)$ reads
\begin{equation}
H_a(t_i)=\frac{1}{t_i}+\frac{t_i}{3}+\frac{23}{180}t_i^3\,.
\end{equation}
The coefficient $c_4(x)$ is thus fixed:
\begin{equation}
c_4(x)=-\frac{3x^2+13x+10}{384}\,.
\end{equation}
Furthermore, the initial condition for the potential $V(t_i)$ is given according to the constraint equation \eqref{constraint}. 

\begin{figure}[t]
\includegraphics[scale=0.55]{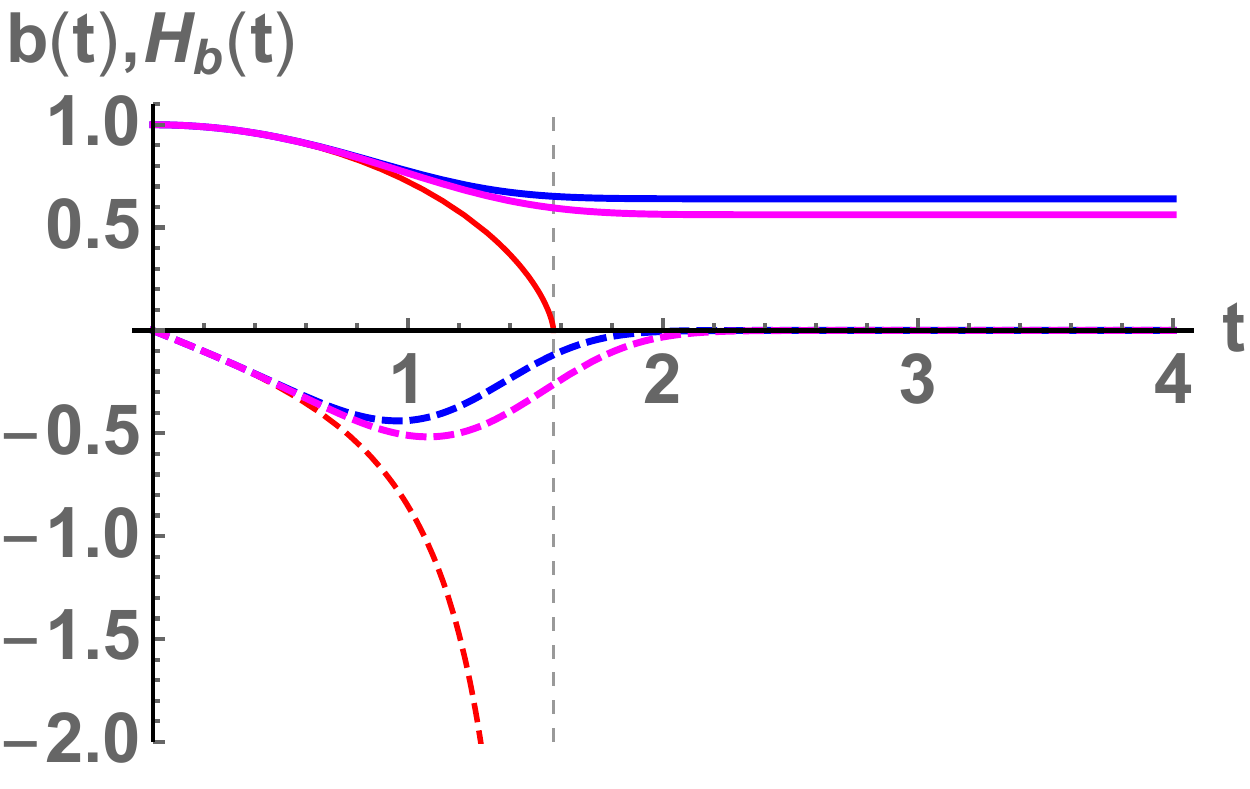}
\includegraphics[scale=0.55]{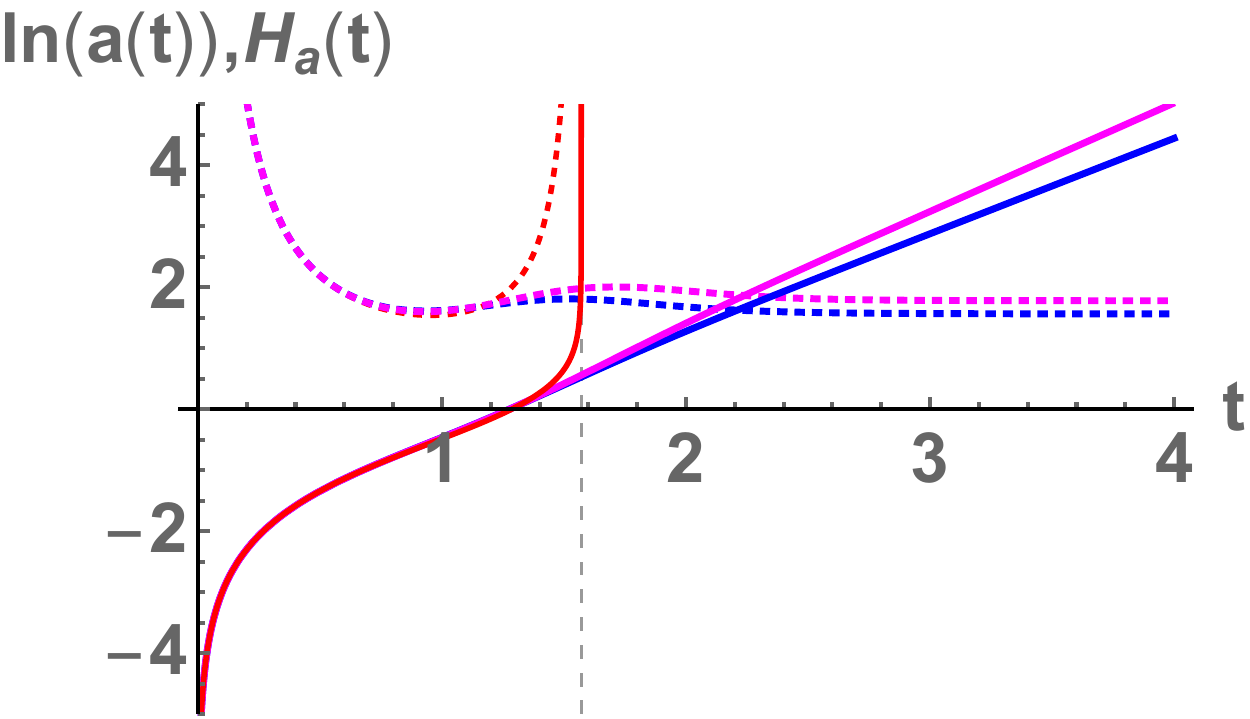}
\includegraphics[scale=0.55]{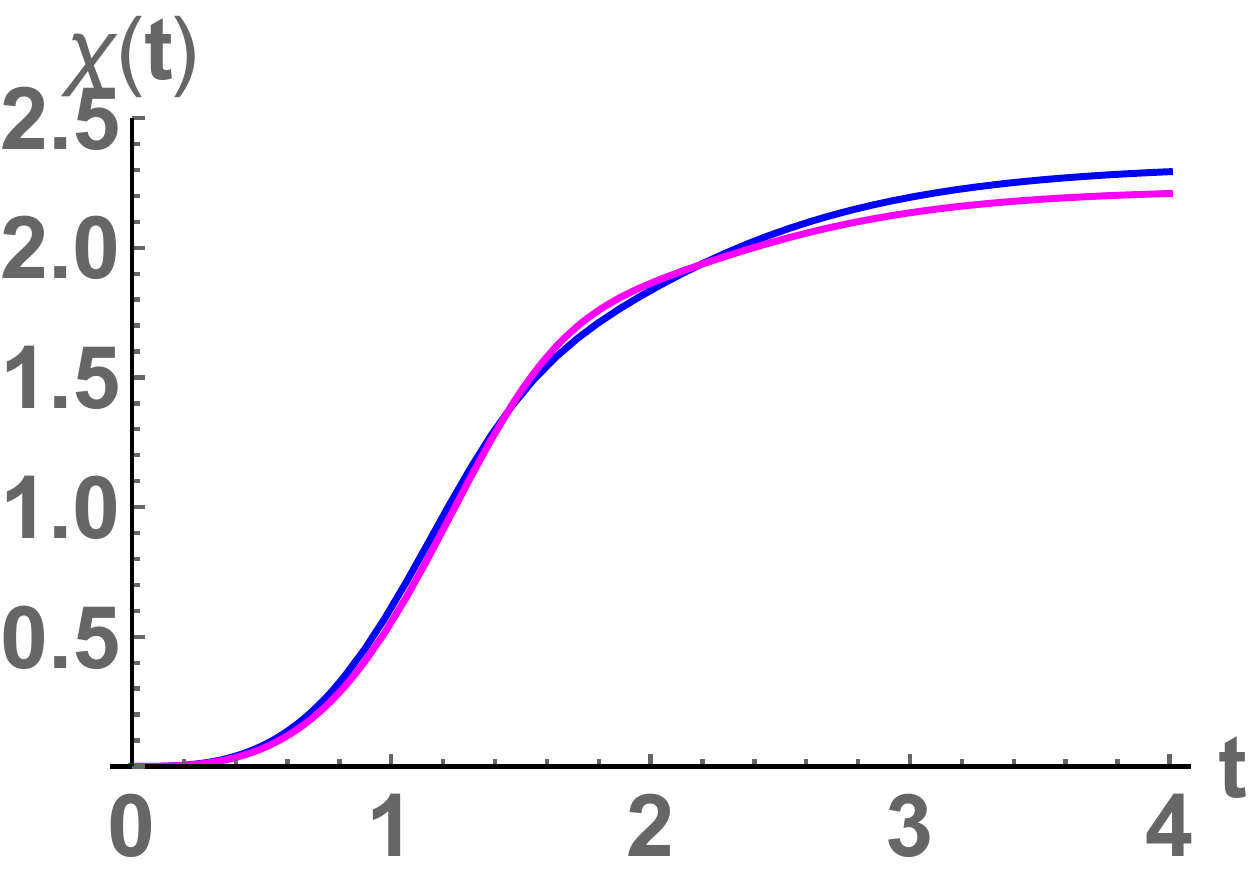}
\includegraphics[scale=0.55]{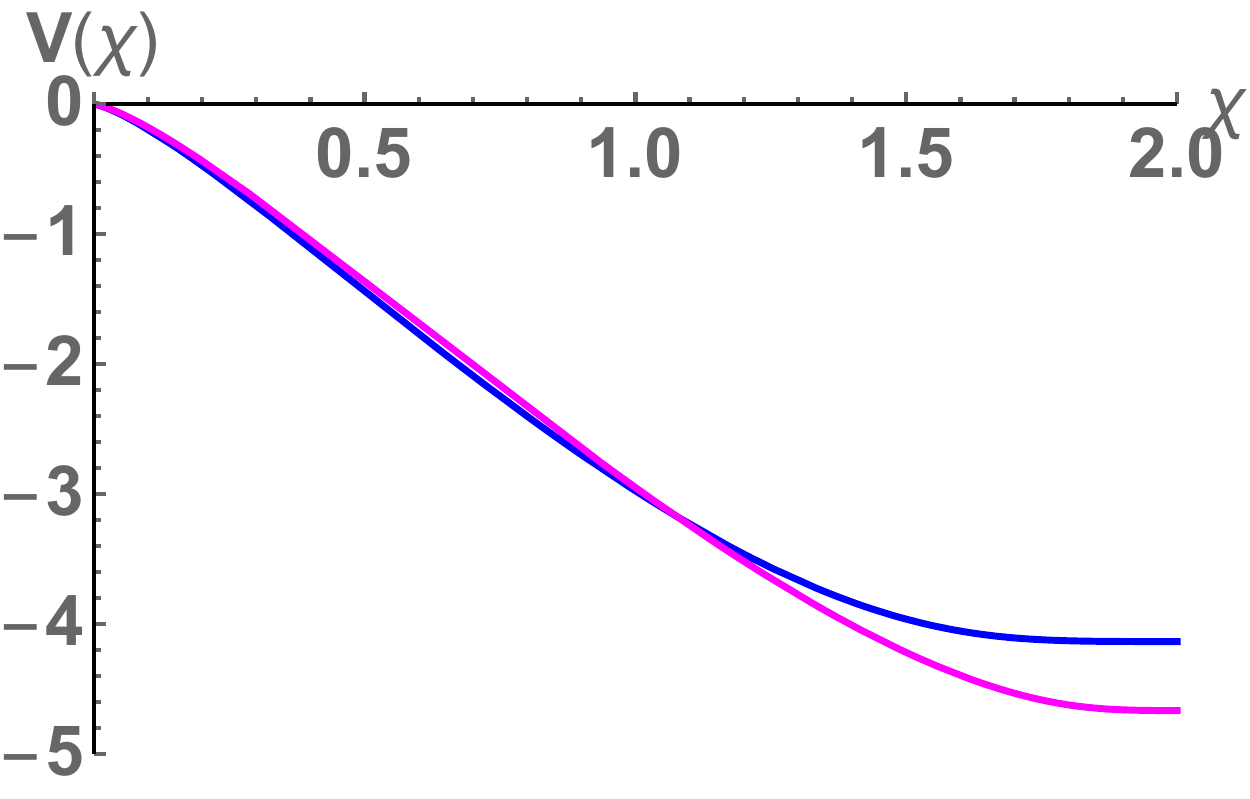}
\caption{\label{solution1}The results of the regular black hole model given by Eq.~\eqref{b1} for $x=4$ (blue) and $x=3$ (magenta), respectively. The red curves are the results of the Schwarzschild spacetime. Top-left: $b(t)$ (solid) and $H_b(t)$ (dashed). Top-right: $\ln a(t)$ (solid) and $H_a(t)$ (dotted). The 3-form field $\chi(t)$ and the potential $V(\chi)$ are shown in the bottom-left and bottom-right panels, respectively. The vertical dashed line ($t=\pi/2$) stands for the singularity in the Schwarzschild black hole. Note that we have rescaled $t/r_s\rightarrow t$. In addition, the gravitational constant is set to $\kappa=1$ in these plots. Likewise for the plots below.}
\end{figure}

We rescale the gravitational constant such that $\kappa=1$ for simplicity. The numerical results are shown in Fig.~\ref{solution1}. The blue and magenta curves represent the results of the regular black hole with $x=4$ and $x=3$, respectively. The red curves correspond to the Schwarzschild interior spacetime, in which the singularity is labeled by the vertical dashed line ($t=\pi/2$). It can be seen that in the regular black hole model, the 3-form field $\chi$ increases in $t$ and approaches its maximum value $\chi_m$, which corresponds to the local minimum $V_m$ of the potential. Also, in the regular black hole model, $H_a(t)$ approaches a constant when $t\rightarrow\infty$. More precisely, $H_a(t)$ and the potential can be approximated as
\begin{equation}
H_a(t)\rightarrow\frac{1}{b_m}\,,\qquad \kappa V\rightarrow\kappa V_m= \frac{1}{b_m^2}\left(1-\frac{\kappa\chi_m^2}{2}\right)\,,\qquad \textrm{when }t\rightarrow\infty\,.\label{approxsolutions}
\end{equation}
Therefore, the scale factor $a(t)$ would be exponentially growing in $t$ when $t\rightarrow\infty$. 

\begin{figure}[t]
\includegraphics[scale=0.55]{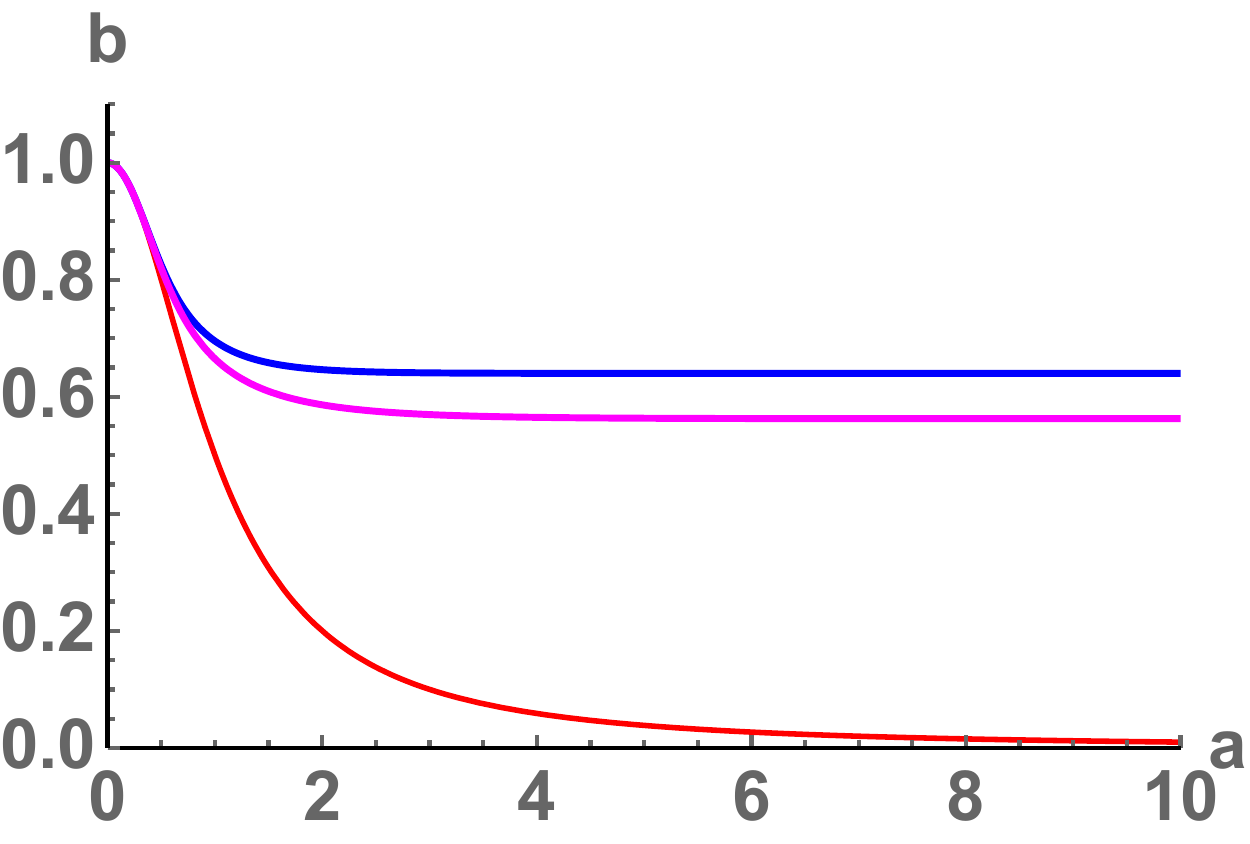}
\caption{\label{solution12c}The scale factor $b$ as a function of $a$ is shown. The magenta and blue curves correspond to $x=3$ and $x=4$ respectively. The red curve shows the result of the Schwarzschild spacetime: $b=1/(1+a^2)$.}
\end{figure}

In fact, one can also express the scale factor $b$ in terms of $a$, such that the result can be presented in a coordinate-independent way. In Fig.~\ref{solution12c}, the scale factor $b$ is shown as a function of $a$ for the regular black hole model with $x=3$ and $x=4$, in magenta and blue, respectviely. Note that the two scale factors of the Schwarzschild metric satisfy $b=1/(1+a^2)$ and they are shown by the red curve.

\begin{figure}[t]
\includegraphics[scale=0.55]{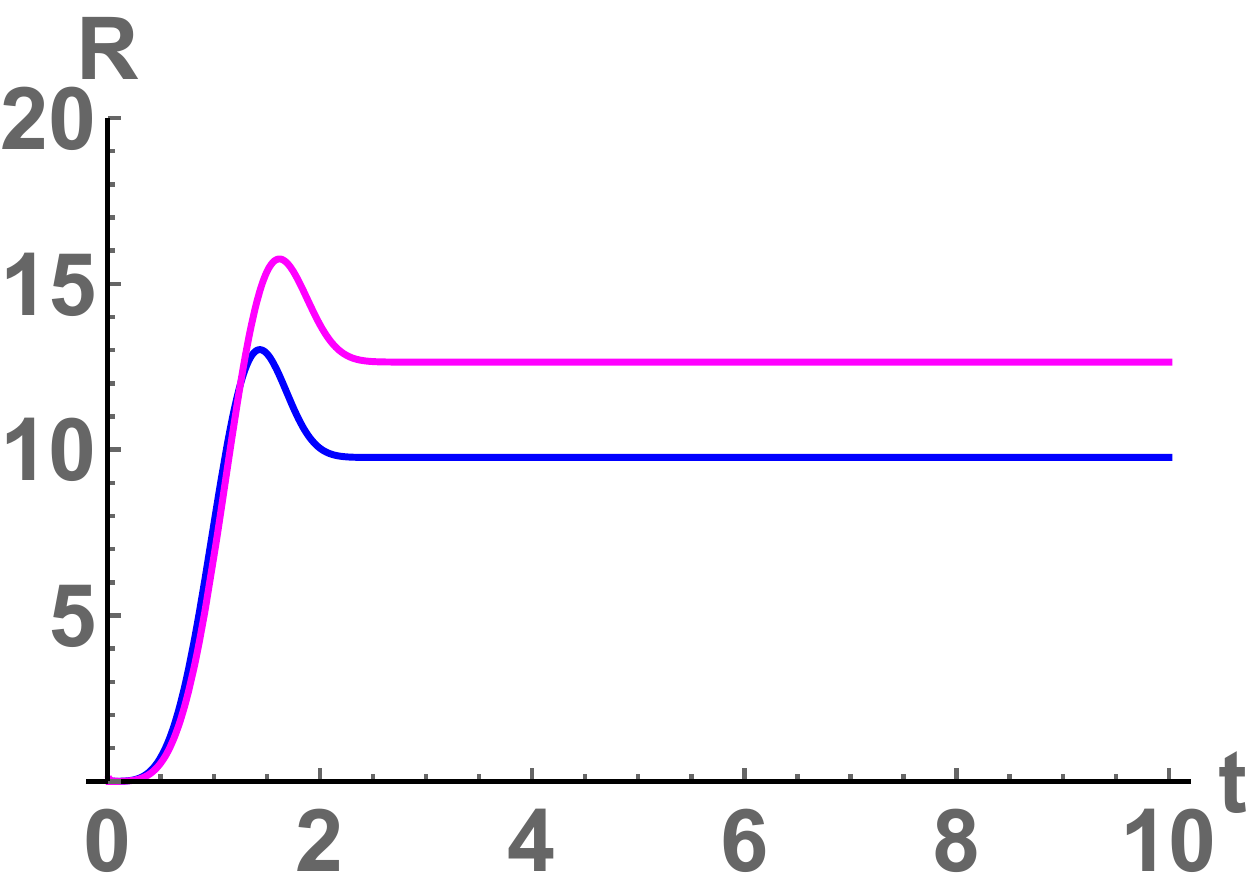}
\includegraphics[scale=0.55]{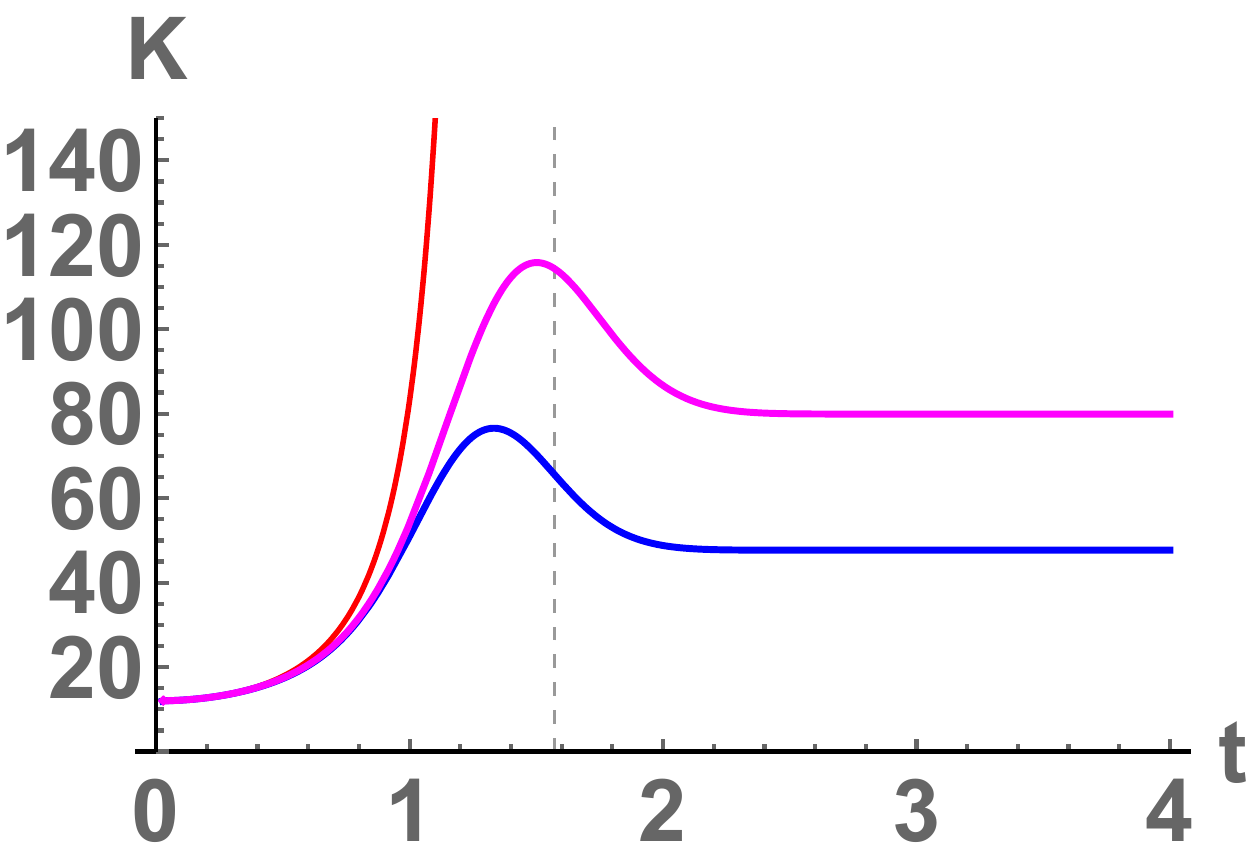}
\caption{\label{solutionR1}The Ricci scalar $R$ and the Kretschmann scalar $K$ of the regular black hole model (the magenta and blue curves correspond to $x=3$ and $x=4$ respectively), compared to the Schwarzschild case in red. The Ricci scalar is of course identically zero in the Schwarzschild case, hence not shown in the plot.}
\end{figure}

In Fig.~\ref{solutionR1}, we show the Ricci scalar $R$ (left) and the Kretschmann scalar $K$ (right) of the regular black hole. Again, the red curve in the right panel shows the Kretschmann scalar of the Schwarzschild black hole ($K_{Sch}=12/b^6$). Note that the Ricci scalar of the Schwarzschild spacetime is identically zero since it is a vacuum solution.

It is surprising that when $b\rightarrow b_m$, the interior spacetime is asymptotically a \emph{2-dimensional} de Sitter spacetime times a 2-sphere, $\text{dS}_2 \times \text{S}^2$. The reason why a $\text{dS}_2$ chart appears accompanied by a negative 3-form potential can be appreciated by writing down the trace of the Einstein equation:
\begin{align}
R=&-\kappa T\nonumber\\
=&\,\kappa\left[2\left(\dot\chi+H_a\chi+2H_b\chi\right)^2+4V-3\chi \frac{dV}{d\chi}\right]\,.\label{RTeq}
\end{align}
Even though the potential is negative, the first term in the second line of Eq.~\eqref{RTeq} also contributes. In the asymptotic limit ($t\rightarrow\infty$), Eq.~\eqref{RTeq} becomes
\begin{equation}
R=\kappa\left(2H_a^2\chi_m^2+4V_m\right)=\frac{4}{b_m^2}\,,
\end{equation}
where we have used Eq.~\eqref{approxsolutions} in the last equality. This result can be interpreted as the 3-form field itself behaving as an effective cosmological constant, which compensates the negative contribution from the potential term. 

\subsection{Spacetime Structure When $b\rightarrow b_m$}

\begin{figure}
\begin{center}
\includegraphics[scale=1.5]{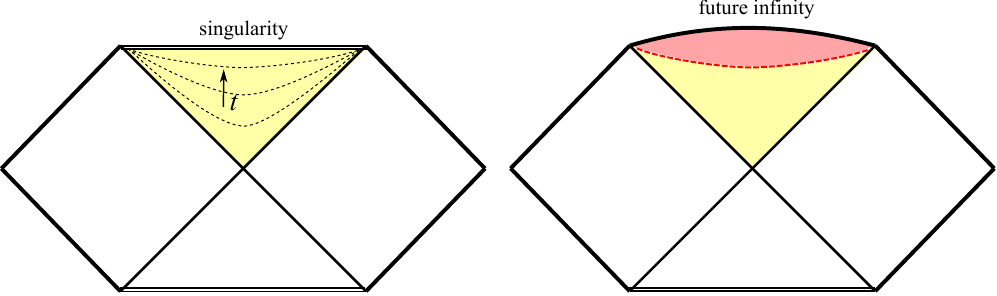}
\caption{\label{fig:causal}Left: The causal structure of a maximally extended Schwarzschild black hole, where our coordinate covers inside the event horizon (yellow colored region), where the time coordinate varies from $t = 0$ (horizon) to $t = \pi/2$ (singularity). Right: The effects of the 3-form field is to modify the solution near the putative singularity. The areal radius approaches a constant and the singularity is replaced by the topology $\text{dS}_2 \times \text{S}^2$ (red colored region). Therefore, one can interpret that the internal structure will evolve to a spacelike future infinity rather than a spacelike singularity.}
\end{center}
\end{figure}

After obtaining the interior geometry of a black hole solution, it is important to investigate the structure of the interior geometry, such as the causal structure of the spacetime. After adopting the following rescalings: $t/r_s\rightarrow t$, $\mathcal{R}/r_s\rightarrow\mathcal{R}$, the metric line element when $b\rightarrow b_m$ can be written as
\begin{equation}
ds^2=-dt^2+e^{2H_at}d\mathcal{R}^2+b_m^2d\Omega_2^2\,.\label{bmmetric}
\end{equation}
Since $H_a\rightarrow 1/b_m$ in this limit, the metric \eqref{bmmetric} is essentially the Nariai solution at large $t$ limit \cite{Nariai1,Nariai2}. In fact, the full Nariai spacetime could be a self-consistent solution in full quantum gravity \cite{Kofman:1985dw}. We would like to emphasize that, while in Refs.~\cite{Bohmer:2007wi,Boehmer:2008fz,Brannlund:2008iw,Dadhich:2015ora}, the authors have shown that similar regular black holes can be formulated within effective models of loop quantum gravity, our result here is obtained in classical general relativity. To scrutinize the causal structure, let us focus on the $t$-$\mathcal{R}$ plane. Introducing a new coordinate $d\bar{r}=e^{-H_at}dt$, the $t$-$\mathcal{R}$ sector of the metric \eqref{bmmetric} reads
\begin{equation}
ds^2=e^{2H_at}\left(d\mathcal{R}^2-d\bar{r}^2\right)\,.
\end{equation}
Then, we introduce a new set of coordinates
\begin{equation}
\bar{u}=e^{\bar{A}(\mathcal{R}+\bar{r})}\,,\qquad\bar{v}=-e^{-\bar{A}(\mathcal{R}-\bar{r})}\,,
\end{equation}
such that
\begin{equation}
d\bar{u}d\bar{v}=\bar{A}^2e^{2\bar{A}\bar{r}}\left(d\mathcal{R}^2-d\bar{r}^2\right)\,,
\end{equation}
where $\bar{A}$ is a positive constant. Finally, we define a set of timelike and spacelike coordinates: $\bar{T}=(\bar{u}-\bar{v})/2$ and $\bar{X}=(\bar{u}+\bar{v})/2$, such that $-d\bar{T}^2+d\bar{X}^2=d\bar{u}d\bar{v}$. The line element becomes
\begin{equation}
ds^2=\bar{A}^{-2}\textrm{exp}\left[2\left(H_at-2\bar{A}\bar{r}\right)\right]\left(-d\bar{T}^2+d\bar{X}^2\right)\,,
\end{equation}
and we have 
\begin{equation}
\bar{T}^2-\bar{X}^2=e^{2\bar{A}\bar{r}}\,.
\end{equation}
When $b\rightarrow b_m$, we have $t\rightarrow\infty$ and $\bar{r}\rightarrow0$. Therefore, the surface $b=b_m$ is a spacelike surface, on which $\bar{T}^2-\bar{X}^2=1$. The causal structure of the interior spacetime can be illustrated by the Penrose diagram, which is shown in Fig.~\ref{fig:causal}. See also the results in Ref.~\cite{Bohmer:2007wi}.

\begin{figure}[t]
\includegraphics[scale=0.55]{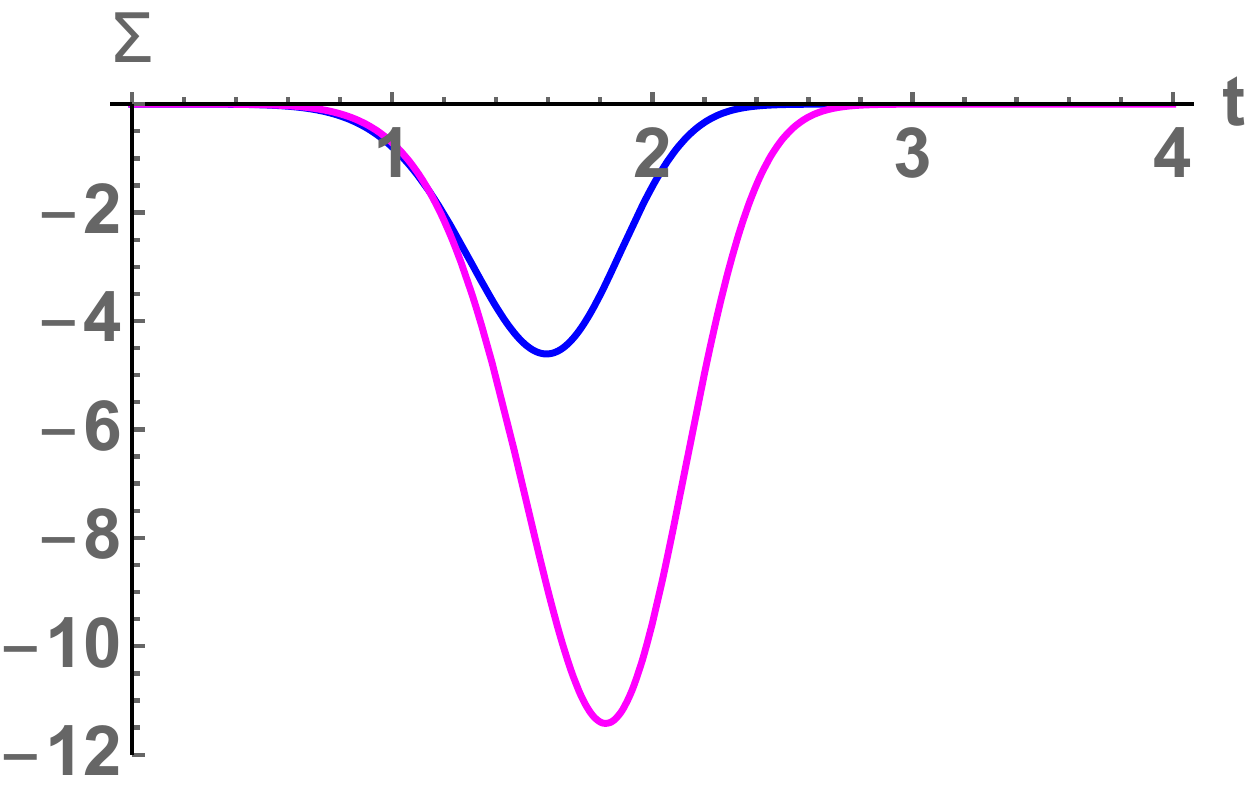}
\caption{\label{solutionNEC}The null energy condition is violated in the regular black hole model ($\Sigma:=T_{\mu\nu}k^\mu k^\nu<0$). The blue and the magenta curves correspond to $x=4$ and $x=3$, respectively.}
\end{figure}

Singularity theorems require various assumptions, such as globally hyperbolicity and energy conditions. Indeed, the reason why the singularity can be avoided in this model is because the null energy condition is violated. Intuitively, the violation of energy conditions effectively gives rise to a ``repulsive force" in the interior spacetime, preventing the formation of spacetime singularities. To see the violation explicitly, we recall that the null energy condition is defined as
\begin{equation}
\Sigma:=T_{\mu\nu}k^\mu k^\nu\geqslant 0\,,
\end{equation}
where $k^\mu$ is a null vector. 
To check that the null energy condition holds, one has to check that $\Sigma \geqslant 0$ for \emph{all} null vectors. On the other hand, to show that it does not hold, it suffices to show a counter-example.
To this aim, let us choose the null vector
\begin{equation}
k^\mu=\left(a(t),1,0,0\right)\,.
\end{equation}
The expression of $\Sigma$ can be written explicitly as
\begin{equation}
\Sigma=a^2\chi\frac{dV}{d\chi}\,.
\end{equation}
Since in our regular black hole model, the 3-form field $\chi$ is positive and $dV/d\chi$ is negative, the null energy condition is therefore violated ($\Sigma<0$). The violation of the null energy condition in this model can be seen in Fig.~\ref{solutionNEC}.

Let us summarize several geometrical properties of the interior spacetime:
\begin{itemize}
\item[--] \textit{Singularity theorem}: In many regular black hole models, there is an inner Cauchy horizon \cite{b1,AyonBeato:1999rg, Hayward:2005gi, Modesto:2005zm, Bronnikov:2005gm,Nicolini:2005vd,FMM2}. That is the reason why several regular black hole models are free from the singularity theorem \cite{Hawking:1973uf}. Due to the existence of the Cauchy horizon, the global hyperbolicity is not satisfied, and therefore, even without violating the null energy condition, a black hole can be free from singularity. However, in our case, there is no Cauchy horizon, but the singularity theorem is violated due to the explicit violation of the null energy condition.
\item[--] \textit{Free from mass inflation}: There is neither an inner Cauchy horizon nor another white hole horizon. The usual problem with the second horizon is the instability issue due to the infinite blue-shift of modes \cite{Poisson:1990eh}; hence, the true geometry must be non-perturbatively investigated and the final results may indicate there exists a curvature singularity \cite{Gursel:1979zza,Gursel:1979zz,Novikov:1980ni,Hong:2008mw,Hwang:2010im}. However, in our example, there is no unstable horizon. This can be a minimal modification of the interior geometry of a black hole which resolves the singularity problem.
\item[--] \textit{Dynamical compactification}: One interesting observation is that the four-dimensional geometry seems to be dynamically compactified into a two-dimensional de Sitter spacetime times a sphere with a constant size. Although this is beyond the scope of this paper, one may further investigate whether there can be a compactification process from higher dimensions to four dimensions. If the 3-form field or another field can realize such a process, this can not only solve the singularity problem, but also the compactification problem of string theory.
\item[--] \textit{De Sitter-like phase from negative potential}: The apparent potential of the 3-form field is negative-definite, but after all arrangements, the interior geometry looks like a de Sitter phase (though with the aforementioned ``compactification''). This might shed some light on the origin of the de Sitter geometry of our Universe, a problem exacerbated by the Swampland conjecture \cite{0509212,1806.08362,1903.06239}.
\end{itemize}

\subsection{Geodesic Equations}

In general relativity, there are two distinct notions of spacetime singularities. The first type is curvature singularities, in which curvature invariants diverge. In the context of the singularity theorems, however, it is a different notion of singularity that appears -- that of geodesic incompleteness. Mathematically, it is possible to have geodesic incompleteness without curvature singularity. For example, one can remove a point by hand in a Minkowski spacetime. However, \emph{physically}, if there is geodesic incompleteness, then one typically suspects that it arises because of strong gravitational field that ``breaks'' the spacetime. Therefore it is expected that geodesic incompleteness should accompany curvature singularity in realistic settings. 

\begin{figure}[t]
\includegraphics[scale=0.7]{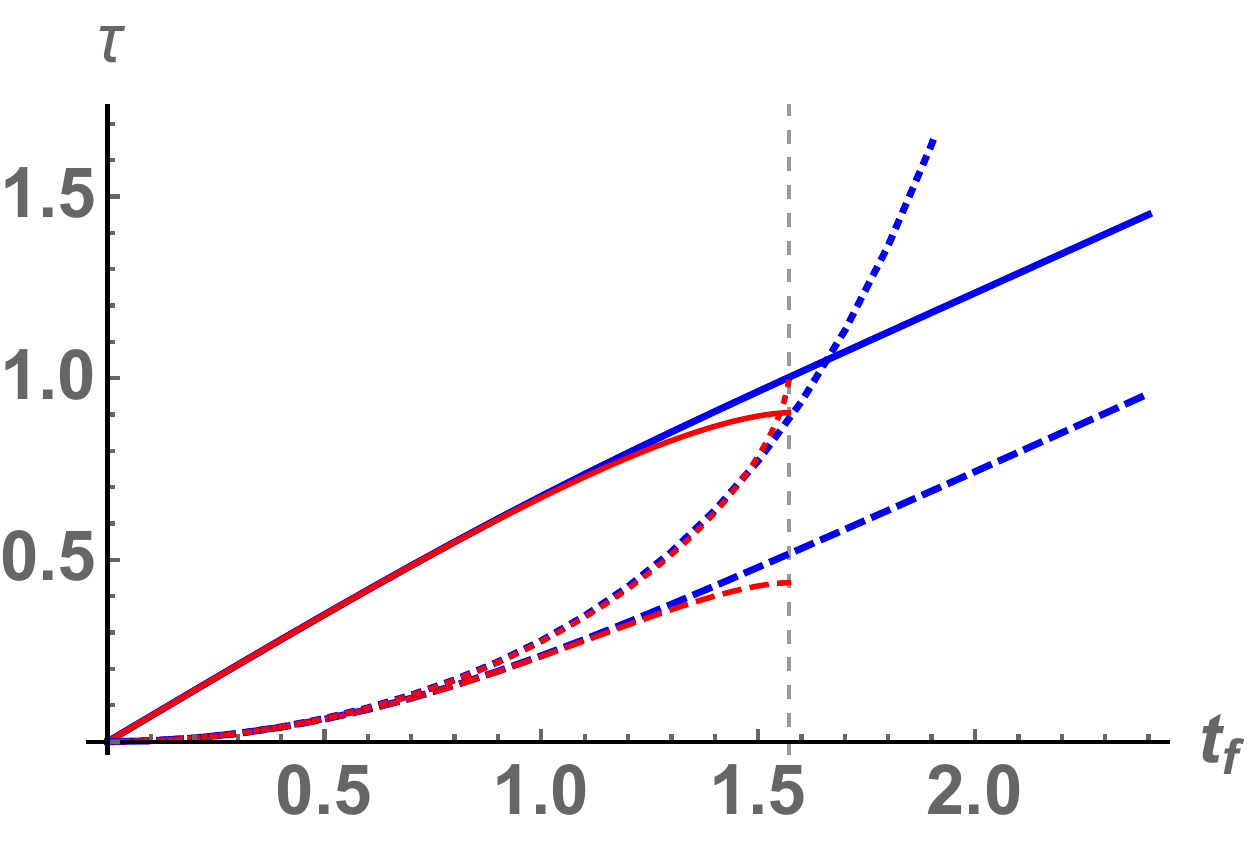}
\caption{\label{fig.pt}The proper time $\tau$ elapsed during the journey from the event horizon $(t=0)$ to a spacelike interior surface labeled by $t=t_f$. The red curves are the results in the Schwarzschild spacetime, in which the singularity is labeled by the vertical dashed line $(t_f=\pi/2)$. The blue curves, on the other hand, are the results in the regular black hole spacetime with $x=4$. Solid, dashed, and dotted curves correspond to $\epsilon=0$; $\epsilon\ne0$ and $\delta+L^2/b_m^2\ne0$;  $\epsilon\ne0$ and $\delta+L^2/b_m^2=0$, respectively.}
\end{figure}

In any case, despite already showing that curvature is bounded in our black hole interior, let us now investigate the properties of geodesic equations and show geodesic completeness explicitly. The geodesic equations of the metric \eqref{KSmetric} contain the following constants of motion:
\begin{equation}
\epsilon\equiv-g_{\mu\nu}u^{\mu}\left(\frac{\partial}{\partial\mathcal{R}}\right)^{\nu}\,,\qquad L\equiv g_{\mu\nu}u^{\mu}\left(\frac{\partial}{\partial\phi}\right)^{\nu}\,,\label{epsilonL}
\end{equation}
where $u$ denotes the 4-velocity of the particle.
$\epsilon$ and $L$ can be interpreted as the conserved energy and the angular momentum along the geodesic, respectively. With the metric \eqref{KSmetric}, the above equations \eqref{epsilonL} can be written as
\begin{equation}
\epsilon=-a(t)^2\frac{d\mathcal{R}}{d\tau}\,,\qquad L=b(t)^2\frac{d\phi}{d\tau}\,.
\end{equation}
Considering the equatorial motion ($\theta=\pi/2$), the geodesic equation reads
\begin{equation}
-\delta=-\left(\frac{dt}{d\tau}\right)^2+\frac{\epsilon^2}{a(t)^2}+\frac{L^2}{b(t)^2}\,,
\end{equation}
where $\delta=0$ and $\delta=1$ corresponds to lightlike and timelike geodesics, respectively. The numerical calculations of the proper time elapsed during the journey from the event horizon ($t=0$) to a spacelike surface inside the black hole ($t=t_f$) is shown in Fig.~\ref{fig.pt}. The blue curves and the red curves are the results in the regular black hole model ($x=4$) and those in the Schwarzschild black hole, respectively. We split the discussions into the following three cases:
\begin{itemize}
\item $\epsilon=0$:
The result is shown by the solid curves ($\epsilon=0$, $\delta=L=1$) in Fig.~\ref{fig.pt}. For the regular black hole model, we consider the limit $b(t)\rightarrow b_m$ and $a(t)\rightarrow e^{H_at}$.  In this limit, the proper time when $b\rightarrow b_m$ can be solved as
\begin{equation}
\tau\approx\frac{t}{\sqrt{\delta+\frac{L^2}{b_m^2}}}\,.\label{approxgeo1}
\end{equation}
Therefore, when $t\rightarrow\infty$, the proper time $\tau$ diverges linearly in $t$.

\item $\epsilon\ne0$ and $\delta+L^2/b_m^2\ne0$: The result is shown by the dashed curves ($\epsilon=\delta=L=1$) in Fig.~\ref{fig.pt}. In the regular black hole model, when $t\rightarrow\infty$, the term $\epsilon^2/a^2$ is negligible compared with the $\delta+L^2/b_m^2$ term. Therefore, the approximated proper time $\tau$ in this limit reduces to Eq.~\eqref{approxgeo1}. When $t\rightarrow\infty$, the proper time $\tau$ diverges linearly in $t$ as well.

\item $\epsilon\ne0$ and $\delta+L^2/b_m^2=0$: The result is shown by the dotted curves ($\epsilon=1$, $\delta=L=0$) in Fig.~\ref{fig.pt}. For the regular black hole, the approximated proper time in the limit $t\rightarrow\infty$ can be solved as
\begin{equation}
\tau\approx\frac{e^{H_at}}{\epsilon H_a}\,.
\end{equation}
Therefore, the proper time $\tau$ diverges exponentially in $t$.

\end{itemize}
We have shown that in the regular black hole model, whenever the particles reach $t\rightarrow\infty$ ($b\rightarrow b_m$), their proper time always diverges. Therefore, particles, no matter massive or massless, would take infinite proper time or affine parameter to reach the surface $b=b_m$. This ensures the geodesic completeness, as well as the regularity of the spacetime. In Ref.~\cite{Carballo-Rubio:2019fnb}, the authors introduced a classification of non-singular black holes based on the behaviors of spacetime geodesics. According to the behavior of geodesic congruences in this model, the asymptotic non-singular geometry belongs to the case B.II in their classification.

Before closing this section, we would like to emphasize that although there is a degree of freedom to tune in the 3-form potential, not all kinds of regular black hole models are attainable in the setup considered in this paper. For example, given the fact that the Kantowski-Sachs metric can only describe the interior cosmological spacetime, this setup is not able to describe regular black holes which contain inner event horizons, such as the model mentioned in Eq.~\eqref{nedmetric}, the self-dual black hole \cite{Modesto:2008im,Hossenfelder:2009fc}, and the regular black holes in the context of non-commutative geometry \cite{Nicolini:2005vd}. The original spacetime singularities in some of these models are replaced with a regular de Sitter core. The Kantowski-Sachs spacetime is not able to completely describe the spacetimes of these models inside the exterior event horizon. In addition, in several effective regular black holes formulated within loop quantum gravity \cite{Ashtekar:2018lag,Ashtekar:2018cay,Bodendorfer:2019cyv,Bodendorfer:2019nvy}, the singularity is replaced with a spacelike transition surface, which connects two asymptotically Schwarzschild spacetimes. One characteristic of this type of ``bouncing" solutions is that the transition surface can be reached by an infalling particle in its finite proper time, which is not the case for the 3-form model obtained in this paper{\footnote{In fact, the behavior of the scale factor $b(t)$ in our case is qualitatively analogous to the loitering effect in cosmology, where the scale factor reaches a minimum value in an infinite cosmic time \cite{Banados:2010ix}.}. In fact, for a bouncing model expressed} by the Kantowski-Sachs metric, the transition surface is characterized by a minimum value of the scale factor $b$, and the other scale factor $a$ reaches its maximum at the transition surface. It should be stressed that these types of regular black holes do not exist in the 3-form theory considered in this paper. This can be directly seen from the equations of motion. More precisely, the transition surface resembles a ``bouncing" property of the scale factors in the sense that
\begin{equation}
H_a=H_b=0\,,\qquad \dot{H}_a<0\,,\qquad \dot{H}_b>0\,,\label{bounce}
\end{equation}
at the transition surface. However, it can be easily seen from Eq.~\eqref{eqc} that such a solution does not exist in the 3-form theory considered in this work, because Eq.~\eqref{eqc} is unlikely to be satisfied at the transition surface.

\section{Discussion}\label{5}

In this work we have shown that 3-form field with an appropriate choice of the potential could support regular black holes whose interior is a cosmological spacetime with topology $\text{dS}_2 \times \text{S}^2$, i.e. a Nariai spacetime. 
Interestingly the radius of  the 2-sphere part is constant (which is governed by the free parameter $x>0$ in the scale factor, Eq. (\ref{b1})), so as the de Sitter part continues to expand exponentially, we have effectively a dynamical compactification -- the universe inside the black hole becomes essentially 2-dimensional at late time. Such kind of dynamical compactification could provide some hints on the origin of the de Sitter geometry of our Universe.

As we have mentioned in the bulk of the paper (Sec.~\ref{4}), another way of obtaining black hole solutions is to start with assuming some particular potentials $V(\chi)$, then solve the field equations to get the corresponding solutions. The authors of Ref.~\cite{Barros:2020ghz} have used this method to obtain spherically symmetric solutions in this theory. Some of the solutions can be interpreted as black holes by identifying the existence of event horizons. The others can be interpreted as naked singularities in which there is no event horizon in the spacetime. However, the analysis regarding the black hole solutions conducted in Ref.~\cite{Barros:2020ghz} only focuses on the exterior region of the spacetime. Since we are interested in formulating regular black hole solutions in this theory, we have mainly focussed on the spacetime inside the event horizon in this work. Therefore, the approach we have followed and the one used in \cite{Barros:2020ghz} can be seen as complementary.

We stress also that our result is completely classical. However, one might postulate some relations with quantum gravity. In a more realistic (and more complicated) model of black hole interior, it is perhaps reasonable to expect that as curvature grows inside a black hole towards the putative singularity, new physics would eventually enter. This might include new fields such as $p$-forms, which would then prevent the singularity from forming, in a similar manner that our simple model is singularity-free. 
One might ask: if there is no singularity, why would there be large curvature in the first place to trigger form fields? The likely answer is that in a more realistic situation one has to take into account the entire dynamical process of gravitational collapse. As matter is crushed into a small region, curvature gets larger. In the standard picture the singularity eventually forms. Our proposal is that the singularity formation is avoided when the 3-form is activated by sufficiently large curvature (sufficiently large energy scales). In other words our current solution that considers a pre-existing black hole is only a first step to check that 3-forms can indeed regularize black hole interior.

Having obtained the regular black hole spacetime, it is natural to ask whether the spacetime is stable against small perturbations, or would it form other singularities through some dynamical collapsing processes? This is beyond the scope of the present paper and we plan to address this issue elsewhere. However, we can still highlight an important point regarding the stabilities of the solution. Unlike many regular black hole solutions, our solution has no inner horizon, and is therefore free from mass inflation instability (which might end in yet another singularity). Although the null energy condition is violated, this is not too much of a concern in a cosmological spacetime \cite{1401.4024}, so it is a small price to pay for resolving the singularity.

Interestingly, it has been proposed that spacetime is fundamentally 2-dimensional at short distances or higher energies \cite{1003.5914, 1009.1136, 1102.3434, 1605.05694,1705.05417}. If the essential features in our simple classical model is representative of what one may find in black hole interior when curvature is sufficiently large, then this provides a dynamical compactification scheme to realize the proposed ``dimensional reduction'' \cite{1009.1136}.

In view of its many applications to cosmology, and now its ability to regularize black hole interior, 3-form field is well-motivated and deserves to be studied more closely for its other utilities in theoretical physics.

\subsection*{Acknowledgement}
The work of M.B.L. is supported by the Basque Foundation of Science Ikerbasque. She
also would like to acknowledge the partial support from the Basque government Grant No. IT956-16 (Spain) and
from the project FIS2017-85076-P (MINECO/AEI/FEDER, UE). She is as well grateful to the kind invitation of the {\textit{Center for Gravitation and Cosmology}} of Yangzhou University where this project was initiated back in December 2019. C.Y.C is supported by Ministry of Science and Technology (MOST), Taiwan, through No. 107-2119-M-002-005 and No. 108-2811-M-002-682. He is also supported by Institute of Physics of Academia Sinica (ASIoP), Leung Center for Cosmology and Particle Astrophysics (LeCosPA) of National Taiwan University, and Taiwan National Center for Theoretical Sciences (NCTS). D.Y and X.Y.C are supported by the National Research Foundation of Korea (Grant No.:  2018R1D1A1B07049126). Y.C.O. thanks NNSFC (grant No.11922508 \& No.11705162) and the Natural Science Foundation of Jiangsu Province (No.BK20170479) for
funding support.

\end{document}